\documentclass[conference]{IEEEtran}

\IEEEoverridecommandlockouts

\usepackage{algorithm}
\usepackage{algorithmic}
\usepackage{amsfonts}
\usepackage{amsthm}
\usepackage{amssymb}
\usepackage{array}
\usepackage{bbm}
\usepackage{bbold}
\usepackage{bm}
\usepackage{booktabs}
\usepackage{caption}
\usepackage{cite}
\usepackage[cmex10]{amsmath}
\usepackage{diagbox}
\usepackage{enumerate}
\usepackage{epstopdf}
\usepackage{footnote}
\usepackage{framed}
\usepackage{graphicx}
\usepackage{indentfirst}
\usepackage{lineno}
\usepackage{multicol}
\usepackage{multirow}
\usepackage{setspace}
\usepackage{subeqnarray}
\usepackage{subfigure}
\usepackage{threeparttable}
\makesavenoteenv{tabular}
\usepackage{url}
\usepackage{xcolor}
\usepackage{setspace}

\newcommand{\mb}{\mathbf}
\newcommand{\mbb}{\mathbb}
\newcommand{\mc}{\mathcal}

\newcommand{\mr}{\mathrm}

\newcommand{\tb}{\textbf}

\newtheorem{theorem}{Theorem}

\begin{document}

	\title{Constrained Error-Correcting Codes for Efficient DNA Synthesis}
		\author
	{
		\IEEEauthorblockN{
		 Yajuan Liu and Tolga M. Duman}\\
		 \IEEEauthorblockA{EEE Department, Bilkent University, Ankara, Turkey\\}
		 Email: yajuan.liu@bilkent.edu.tr, duman@ee.bilkent.edu.tr
		
		\thanks{This work was funded by the European Union through the ERC Advanced
			Grant 101054904: TRANCIDS. Views and opinions expressed are, however,
			those of the authors only and do not necessarily reflect those of the European
			Union or the European Research Council Executive Agency. Neither the
			European Union nor the granting authority can be held responsible for them.}
		}
	
	\maketitle
	
	\begin{abstract}
		DNA synthesis is considered as one of the most expensive components in current DNA storage systems.
		In this paper, focusing on a common synthesis machine, which generates multiple DNA strands in parallel following a fixed supersequence, we propose constrained codes with polynomial-time encoding and decoding algorithms. Compared to the existing works, our codes simultaneously satisfy both $\ell$-runlength limited and $\epsilon$-balanced constraints. By enumerating all valid sequences, our codes achieve the maximum rate, matching the capacity. Additionally, we design constrained error-correcting codes capable of correcting one insertion or deletion in the obtained DNA sequence while still adhering to the constraints.
	\end{abstract}
	
	\section{Introduction}

	DNA storage systems have attracted significant attention  in recent years owing to their remarkable durability and high storage density \cite{church2012next,goldman2013towards,yazdi2015dna}.
	These systems encode digital information into four DNA nucleotides, namely Adenine (A), Thymine (T), Cytosine (C) and Guanine (G).  
	A typical DNA storage system consists of three key stages: synthesis, storage, and sequencing of DNA sequences \cite{laver2015assessing}. In particular, DNA synthesis is regarded as one of the most costly components in current DNA storage systems	\cite{lenz2020coding}.

In the DNA synthesis process, a large number of DNA strands are commonly created  in parallel \cite{caruthers2013chemical,ceze2019molecular,kosuri2014large}.
Generally,
the synthesis machine follows a fixed template supersequence $S$  and adds one nucleotide at a time to form the specific DNA strands. The synthesis process continues until the entire supersequence is exhausted.
 Consequently, 
 the length of the fixed supersequence dictates
 the number of cycles required to generate all the strands, corresponding to the total \textit{synthesis time}. 
 This leads to an interesting constrained coding problem, first introduced by Lenz \textit{et al.} \cite{lenz2020coding} and subsequently studied in \cite{sini2023dna,makarychev2022batch,elishco2023optimal,immink2024constructions,chrisnata2023deletion,nguyen2024efficient,lu2024coding}. 
 In \cite{lenz2020coding}, the authors demonstrate that the supersequences that maximize the information rate are the alternating quaternary sequences. Therefore, in this paper, we employ the alternating sequence $S=ATCGATCG\cdots$ as the fixed template supersequence for DNA synthesis. 
 Moreover, DNA synthesis is prone to various types of errors, such as insertions, deletions, and substitutions. To mitigate those errors, two constraints are typically employed in practice. One is the \textit{$\ell$-runlength limit ($\ell$-RLL)}  \cite{yakovchuk2006base,ross2013characterizing}, which restricts the length of consecutive identical nucleotides in a DNA sequence to be no greater than $\ell$, where $\ell$ is a positive integer. The other is the \textit{$\epsilon$-balance} \cite{ross2013characterizing,heckel2019acharacterization}, which ensures that the ratio of $G$ and $C$ in a DNA sequence lies within the range $[0.5-\epsilon, 0.5+\epsilon]$, where $\epsilon\in[0,0.5]$. 
 To effectively reduce or correct errors in DNA synthesis, it is critical to take into account the constructions of constrained codes that satisfy $\ell$-RLL and/or $\epsilon$-balanced constraints, as well as  constrained  error-correcting codes (ECCs) capable of correcting errors while adhering to these two constraints.
 
	\begin{table*}[t]
	\begin{center}
		\caption{Comparison of different constrained ECCs for DNA synthesis, where $r_*,*\in \{\ell, T, (\ell,T),(\ell,\epsilon, T)\}$ denotes the redundancy of enumeration methods.
		}
		\label{tab:compare}
		\begin{tabular}{ccccc}
			\toprule
			Synthesis time $T$  & Redundancy & Computational Complexity & Property & Ref.   \\
			\midrule
			$T\le 2.5n$  & $1$ & $O(n)$& / &  \cite{lenz2020coding}\\
			$T\le 2.5n$  & $\lceil\log_4 n\rceil +5$ &$O(n)$ & indel &  \cite{chrisnata2023deletion}\\
			$T\le 2.5n - 1$  & $\lceil\log_4 (n-2)\rceil +3$ & $O(n)$&indel & \cite{nguyen2024efficient} \\
			$T\le 2.5n - 1$  & $r_{\ell}+3\lceil\log_4 n\rceil +O(1)$ & $O(n^2\ell^2)$&$\ell$-RLL, indel & \cite{nguyen2024efficient} \\
			$n\le T\le 4n$  & $r_T+\lceil\log_4 n\rceil +3$ & $O(n^3)$&indel & \cite{chrisnata2023deletion}\\
			$n\le T\le 4n$ & $r_T+\lceil\log_4 n\rceil +1$ & $O(n^5)$&indel & \cite{chrisnata2023deletion}\\ 
			$n\le T\le 4n$  & $r_{\ell,T}$ & $O(n^3\ell^2)$ & $\ell$-RLL & Thm. \ref{thm:ell}  \\
			$n\le T\le 4n$  & $r_{\ell,\epsilon,T}$ & $O(n^4\ell^2)$ & $\ell$-RLL, $\epsilon$-balance & Thm. \ref{thm:ell,epsilon}\\
			$n\le T\le 4n$  & $r_{\ell,\epsilon,T}+2\lceil\log_4 n\rceil +8$ & $O(n^4\ell^2)$ & $\ell$-RLL, $\epsilon$-balance, indel & Thm. \ref{thm:ell,epsilon,indel} \\
			\bottomrule
		\end{tabular}
	\end{center}
\end{table*}

Building upon the advancements in the DNA synthesis machine, various studies have explored both the theoretical aspects and code constructions for DNA synthesis \cite{sini2023dna, makarychev2022batch, elishco2023optimal, immink2024constructions,chrisnata2023deletion, nguyen2024efficient, lu2024coding}.
Specifically,
 Sini \textit{et al.} \cite{sini2023dna} investigated optimizing DNA synthesis by using shortmers instead of single nucleotides. Makarychev \textit{et al.} \cite{makarychev2022batch} studied batch optimization techniques to reduce the cost of large-scale DNA synthesis.
 Building on this work, Elishco \textit{et al.} \cite{elishco2023optimal} advanced the theoretical understanding of DNA synthesis by modeling the process as a constrained system and analyzing the $\ell$-RLL constraint. Immink \textit{et al.} \cite{immink2024constructions} explored the trade-offs between several components required to synthesize multiple DNA strands in parallel.

We can observe that the synthesis time of an arbitrary DNA sequence  with length $n$ is at most $T\le 4n$ during  DNA synthesis. 
Recent developments have demonstrated that introducing redundancy can significantly reduce the synthesis time. For instance, Lenz \textit{et al.} \cite{lenz2020coding} proposed an explicit construction that limits the synthesis time to $T \le 2.5n$ for a  DNA sequence with length $n$, using only one bit of redundancy. Chrisnata \textit{et al.} \cite{chrisnata2023deletion} further enhanced the process by designing systematic ECCs that can correct single insertion/deletion (indel) error, while maintaining the same synthesis time of $T \le 2.5n$, but with $\lceil \log_4 n \rceil + 5$ symbols of redundancy. 
Recently, Nguyen \textit{et al.} \cite{nguyen2024efficient} have developed constrained ECCs that achieve a synthesis time of $T \le 2.5n - 1$ while ensuring the $\ell$-RLL constraint. The results from \cite{lu2024coding} demonstrate two families of codes capable of correcting synthesis defects.
Additionally,  many other studies have focused on constrained codes with $\ell$-RLL or/and $\epsilon$-balance for DNA storage systems \cite{liu2022capacity,song2018codes,van2010construction, nguyen2021capacity,he2024efficient}. However, these works have predominantly concentrated on DNA storage systems as a whole, rather than DNA synthesis. Thus, the challenge of constructing constrained ECCs that simultaneously satisfy both the $\ell$-RLL and $\epsilon$-balance constraints ($(\ell, \epsilon)$-constraints) remains an open problem in the context of DNA synthesis.

  To address this issue,  we propose explicit constrained codes that incorporate both $\ell$-RLL and $\epsilon$-balance for DNA synthesis, with a fixed synthesis time $T$ satisfying $n\le T\le 4n$. These codes encode the binary information into the DNA sequences with polynomial-time encoding and decoding algorithms. By enumerating all valid sequences, the rate of our codes achieves the theoretical capacity.
  Furthermore, we develop constrained ECCs for DNA synthesis, which can correct one indel in the DNA strand.
  We compare the designed constructions with some existing results in Table \ref{tab:compare}.

	The rest of this paper is organized as follows. In Section \ref{sec:pre}, some necessary preliminaries are introduced. Then the constrained codes with $\ell$-RLL or $(\ell,\epsilon)$-constraints are developed in Section \ref{sec:constrained-code}. In Section \ref{sec:constrained-error-code}, we construct  constrained ECCs to correct one indel for DNA synthesis. Finally, the concluding remarks are drawn in Section \ref{sec:conclusion}.
	
	\section{Preliminaries}\label{sec:pre}
	
	Let $\Sigma_4=\{1,2,3,4\}$. For a DNA system, we map the four different bases as follows: $A\rightarrow 1,T\rightarrow 2,C\rightarrow 3$ and $G\rightarrow 4$, and use the shifted modulo operator $(a\bmod 4) \in\{1, 2, 3, 4\}$ throughout this
	paper. Denote $[a,b]=\{a,a+1,\dots,b\}$ and  $[a,b)=\{a,a+1,\dots,b-1\}$ as two order sets for any two non-negative integers $a,b$ with $a<b$,  respectively.
	Let $\{a\}^t$ be a length-$t$ sequence formed by $a \in \Sigma_4$.
	For a DNA sequence $\mathbf{c} \in \Sigma_4^n$, we write
	$$\mathbf{c}=(\{a_1\}^{t_1}||\{a_2\}^{t_2}||\cdots|| \{a_k\}^{t_k}),a_i\in \Sigma_4,~t_i \in [1, n],$$
	where $a_i\ne a_{i+1},i\in [1,k),\sum_{i=1}^kt_i=n$ and $||$ denotes a concatenation of two sequences (e.g., $\{a_1\}||\{a_2\}^2||\{a_3\}=a_1a_2a_2a_3$). Particularly, $\mathbf{c}|| \emptyset=\mathbf{c}$.
	Each  sub-sequence  $\{a_i\}^{t_i}$ is called a \textit{run} in $\mathbf{c}$, and $\{a_k\}^{t_k}$ is the \textit{last run}.
	We say that $\mathbf{c}$ satisfies \textit{$\ell$-RLL} if $t_i \in [1, \ell]$, $\forall ~i \in [1, k]$. In this paper, we consider the set of length-$n$ DNA sequences with $\ell$-RLL, thus, all valid runs are included in the following set,
	\begin{eqnarray}\label{set:S}
		\mathcal{S}=\left\{\{a\}^t:a \in \Sigma_4,~t \in [1, \ell]\right\}.
	\end{eqnarray}
	
	For a run $\mathbf{s}=\{a\}^t \in\mathcal{S} $, denote the symbol and the length of $\mathbf{s}$ by $e(\mathbf{s})=a$ and $|\mathbf{s}|=t$, respectively.
	For a sequence $\mathbf{c}=(c_1,c_2,\dots, c_n)\in \Sigma_4^n$, denote the  \textit{GC-content} and \textit{synthesis time} of $\mathbf{c}$ by $G(\mathbf{c})$ and $T(\mathbf{c})$, respectively, where
	$$G(\mathbf{c})=\sum_{i=1}^n\mathbb{1}_{\{c_i>2\}},$$
	therein $\mathbb{1}_{\{*\}}$ equals $1$ when condition $*$ is true, otherwise it is $0$.
	We say that $\mathbf{c}$ satisfies \textit{$\epsilon$-balance} if 
	\begin{eqnarray}\label{eqn:M}
	G(\mathbf{c})\in \mathcal{M}=[\lceil{(0.5-\epsilon)n}\rceil, \lfloor{(0.5+\epsilon)n}\rfloor].
	\end{eqnarray}
	 Particularly,  when $n$ is even and $G(\mathbf{c})=0.5n$, i.e., $\epsilon=0$, we say that $\mathbf{c}$ is balanced.
	
		For a DNA sequence $\mathbf{c}=(c_1,c_2,\dots,c_n)$, denote the prefix of $\mathbf{c}$ with length $k$ by $P_{k}(\mathbf{c})=(c_1,c_2,\dots,c_{k}),k\in[1,n]$. 
	Define the \textit{differential sequence} of $\mathbf{c}$ as
	\begin{eqnarray*}
		d(\mathbf{c})&=&\{d_1(\mathbf{c}),d_2(\mathbf{c}),\dots,d_n(\mathbf{c})\}\\
		&\triangleq&\{c_1,(c_2-c_1)\bmod 4,(c_3-c_2)\bmod4,\dots,\\
		&&~~(c_n-c_{n-1})\bmod 4\}.
		\end{eqnarray*}	
	Then, from \cite{lenz2020coding}, the \textit{synthesis time} of $\mathbf{c}$ is  $T(\mathbf{c})=\sum_{i=1}^{n}d_i(\mathbf{c})$.

	For a subset of DNA sequences $\mathcal{C}\subseteq\Sigma_4^n$, we define $\mathcal{C}$ as a \textit{constrained code} if any codeword $\mathbf{c}\in\mathcal{C}$ satisfies the required constraints. Let $|\mathcal{C}|$ be the cardinality of $\mathcal{C}$, then the rate of $\mathcal{C}$ is defined as $(\log_2|\mathcal{C}|)/n$,
	which represents the number of bits stored per nucleotide.
	If $\mathcal{C}$ contains all available codewords under the given constraints, the rate of $\mathcal{C}$ is defined as the \textit{capacity} and $\mathcal{C}$ is said to be \textit{capacity-achieving}.
	For any $\mathbf{c}\in\mathcal{C}$, denote $\mathcal{B}^{DI}(\mathbf{c})$ is the set of sequences obtained from $\mathbf{c}$ by  deleting or inserting one symbol.
	For any $\mathbf{c}_1,\mathbf{c}_2\in\mathcal{C}$, if 
	\begin{eqnarray*}
		\mathcal{B}^{DI}(\mathbf{c}_1)\cap \mathcal{B}^{DI}(\mathbf{c}_2)=\emptyset,
	\end{eqnarray*}	
	then $\mathcal{C}$ is regarded as a \textit{constrained ECC code}.

	\section{Constrained Codes for DNA Synthesis}\label{sec:constrained-code} 
	In this section, we construct  capacity-achieving $\ell$-RLL  and $(\ell,\epsilon)$-constrained codes for DNA synthesis, denoted by $\widetilde{\mathcal{C}}(n,\ell,T)$ and $\widetilde{\mathcal{C}}(n,\ell,\epsilon,T)$, respectively, where $n$ and $T$ are the code length and synthesis time. 
	We begin by partitioning the set of DNA sequences that satisfy the $\ell$-RLL constraint into some subsets, which are subsequently ordered according to specific rules. Following this, we develop unranking and ranking methods to facilitate the encoding and decoding processes, respectively. Both the encoding and decoding algorithms of the developed codes have polynomial-time computational complexities.

	\subsection{$\ell$-RLL Constrained Codes for DNA Synthesis}\label{sec:l-indel}

		For a fixed $T$ with $n\le T\le4n$, denote the set of length-$n$ sequences satisfying the $\ell$-RLL constraint with synthesis time at most $T$  by $\widetilde{\mathcal{C}}(n,\ell,T)$.
	Denote all the DNA sequences with length $n$, last run $\mathbf{s}$ and synthesis time at most $T$ by $\mathcal{C}(n,\mathbf{s},T)$.
	Thus, we can write
	\begin{eqnarray*}
		\widetilde{\mathcal{C}}(n,\ell,T)=\bigcup_{\mathbf{s}\in\mathcal{S}}\mathcal{C}(n,\mathbf{s},T),
	\end{eqnarray*}
where $\mathcal{S}$ is the set of all valid runs defined in \eqref{set:S}.

	Let $N(n,\mathbf{s},T)$ be the size of $\mathcal{C}(n,\mathbf{s},T)$, i.e., $N(n,\mathbf{s},T)=|\mathcal{C}(n,\mathbf{s},T)|$. In the following, we obtain a recursion for computing $N(n,\mathbf{s},T)$.
	
	\begin{theorem}\label{thm:state-ell}
		The number of  length-$n$ DNA sequences with $\ell$-RLL and synthesis time at most $T$ can be computed by 
			\begin{eqnarray}\label{eqn:state-ell}
			&\hspace{-20em} 	N(n,\mathbf{s},T)\nonumber\\
			&=\left\{\begin{array}{ll}
					0 ,  \text{~~~~~~if~}  n<|\mathbf{s}| \text{~or~} T<4(|\mathbf{s}|-1)+e(\mathbf{s}),\\
					\mbb{1}_{\{T\ge4(|\mathbf{s}|-1)+e(\mathbf{s})\}},\qquad\qquad \text{~~~~~if~}  n=|\mathbf{s}|, \\
					\sum\limits_{\mathbf{s}'\in \mathcal{S}:e(\mathbf{s}')\ne e(\mathbf{s})} N(n-|\mathbf{s}|,\mathbf{s}',T'),  \\	
					\qquad\text{~if~}  n>|\mathbf{s}| \text{~and~} T\ge 4(|\mathbf{s}|-1)+e(\mathbf{s}),
				\end{array}\right.
			\end{eqnarray}
		where $\mathbf{s}\in \mathcal{S},T'=T-4(|\mathbf{s}|-1)-(e(\mathbf{s})-e(\mathbf{s}')\bmod 4)$.
	\end{theorem}
	\begin{proof}
		When $n<|\mathbf{s}|$ or $T<4(|\mathbf{s}|-1)+e(\mathbf{s})$, there does not exist a sequence which satisfies $\ell$-RLL and synthesis time at most $T$, resulting in $N(n,\mathbf{s},T)=0$.
		When $n=|\mathbf{s}|$, if one sequence satisfies $\ell$-RLL in DNA synthesis, then the synthesis time must be $T=4(|\mathbf{s}|-1)+e(\mathbf{s})$.

		When $n>|\mathbf{s}|$, for a DNA sequence $\mathbf{c}=(c_1,c_2,\dots,c_n)\in\mathcal{C}(n,\mathbf{s},T)$, the prefix of $\mathbf{c}$ with length $n-|\mathbf{s}|$ is $P_{n-|\mathbf{s}|}(\mathbf{c})=(c_1,c_2,\dots,c_{n-|\mathbf{s}|})$. Denote the last run of $P_{n-|\mathbf{s}|}(\mathbf{c})$ as $\mathbf{s}'$, that is,
		\begin{eqnarray*}
			\mathbf{c}=P_{n-|\mathbf{s}|}(\mathbf{c})||\mathbf{s}=(c_1,c_2,\dots,c_{n-|\mathbf{s}|-|\mathbf{s}'|})||\mathbf{s'}||\mathbf{s},
		\end{eqnarray*}
		where $\mathbf{s}'\in \mathcal{S}, e(\mathbf{s})\ne e(\mathbf{s}')$.
		Since $T(\mathbf{c})\le T$, we know
		\begin{eqnarray*}
			T(P(\mathbf{c}))\le T'=T-4(|\mathbf{s}|-1)-(e(\mathbf{s})-e(\mathbf{s}')\bmod 4).
		\end{eqnarray*}
		
		Therefore, we obtain
	\begin{eqnarray*}
				\mathcal{C}(n,\mathbf{s},T)=\{\mathbf{c}'||\mathbf{s}:
				\mathbf{c}'\in\mathcal{C}(n-|\mathbf{s}|,\mathbf{s}',T'),
				\mathbf{s}'\in \mathcal{S} \\\text{~with~}e(\mathbf{s}')\ne e(\mathbf{s})\},
	\end{eqnarray*}
 concluding the proof.
	\end{proof}

	For any pairs of DNA sequences $\mathbf{c}_1\in \mathcal{C}(n,\mathbf{s}_1,T_1)$ and $ \mathbf{c}_2\in \mathcal{C}(n,\mathbf{s}_2,T_2)$, where $\mathbf{s}_1,\mathbf{s}_2\in\mathcal{S}$,
	we order the sequences logically by the length and symbol of the last run. 
	Define $\mathbf{c}_1< \mathbf{c}_2$ if any of the following conditions are satisfied. 
	\begin{itemize}
		\item[a.] $|\mathbf{s}_1|<|\mathbf{s}_2|$; 
		\item[b.] $|\mathbf{s}_1|=|\mathbf{s}_2|$ and $e(\mathbf{s}_1)<e(\mathbf{s}_2)$; 
		\item[c.] $|\mathbf{s}_1|=|\mathbf{s}_2|,e(\mathbf{s}_1)=e(\mathbf{s}_2)$ and $P_{n_1}(\mathbf{c}_1)<P_{n_2}(\mathbf{c}_2)$, where $n_1=n-|\mathbf{s}_1|,n_2=n-|\mathbf{s}_2|$.
	\end{itemize}
	
After computing the size of each subset and logically ordering all DNA sequences with $\ell$-RLL, we design the encoding and decoding algorithms by unranking and ranking methods, denoted by $\mathrm{Enc}_{\ell,T}$ and $\mathrm{Dec}_{\ell,T}$, respectively. The details are provided in Algorithms \ref{Unrank} and \ref{Rank}.
For the encoding algorithm, we unrank an integer $M\in[1,N(n,\ell,T)]$ to a DNA sequence in $\widetilde{\mathcal{C}}(n,\ell,T)$. For the decoding algorithm, we perform the inverse operation to rank the DNA sequence in $\widetilde{\mathcal{C}}(n,\ell,T)$ to an integer $M\in[1,N(n,\ell,T)]$.	
	
	\begin{algorithm}[htbp]
		\caption{$\mathrm{Enc}_{\ell,T}$: Unrank the $M$-th sequence in $\widetilde{\mathcal{C}}(n,\ell,T)$.}
		\label{Unrank}
		\begin{algorithmic}[1]
			\REQUIRE    $n,  \ell, T,M \in [1, N(n, \ell,T)]$.
			\ENSURE     The $M$-th smallest sequence in $\widetilde{\mathcal{C}}(n,\ell,T)$.
			\RETURN $\mr{unrank}(M, n, \mc{S},T)$.
			\STATE
			
			\STATE  \tb{Function:} $\mr{unrank}(M, n, \mc{S},T)$ \label{line1-4}
			\IF {$n == 0$}
			\RETURN $\emptyset$.
			\ELSE
			\STATE  Find the unique $\mb{s}$ such that $M_1 < M \leq M_1 + N(n, \mb{s}, T)$,
			where 
			\begin{eqnarray*}
				M_1 = \sum_{\mb{s}' \in \mc{S}: e(\mb{s}')< e(\mb{s}) }N(n, \mb{s}',T).
			\end{eqnarray*}\label{line7}
			\STATE  $T'=T-4(|\mathbf{s}|-1)-(e(\mathbf{s})-e(\mathbf{s}')\bmod 4)$.
			\STATE  $\mc{S}' = \mc{S} \setminus \{\{e(\mb{s})\}^t: t \in [1, \ell]\}$.
			\RETURN $\mr{unrank}(M - M_1, n - |\mb{s}|, \mc{S}',T') || \mb{s}$.
			\ENDIF
			
		\end{algorithmic}
	\end{algorithm}
	\vspace{-0.5cm}
	\begin{algorithm}[htbp]
		
		\caption{$\mathrm{Dec}_{\ell,T}$: Rank $\mb{c}$ in $\widetilde{\mathcal{C}}(n,\ell,T)$.}
		\label{Rank}
		\begin{algorithmic}[1]
			\REQUIRE    $n, \ell, T,\mb{c} \in \widetilde{\mathcal{C}}(n,\ell,T)$.
			\ENSURE     The ranking of $\mb{c}$ in $\widetilde{\mathcal{C}}(n, \ell,T)$.
			\RETURN $\mr{rank}(\mb{c}, n,  \mc{S},T)$.
			\STATE
			\STATE  \tb{Function:} $\mr{rank}(\mb{c}, n,  \mc{S},T)$ \label{line2-4}
			\IF {$n == 0$}
			\RETURN $1$.
			\ELSE
			\STATE  Let $\mb{s}$ be the last run of $\mb{c}$.
			\STATE  $M_1 = \sum_{\mb{s}' \in  \mc{S}: e(\mb{s}') < e(\mb{s})} N(n, \mb{s}', T).$  \label{line2-8}
			\STATE  $T'=T-4(|\mathbf{s}|-1)-(e(\mathbf{s})-e(\mathbf{s}')\bmod 4)$.
			\STATE  $\mc{S}' = \mc{S} \setminus \{\{e(\mb{s})\}^t: t \in [1, \ell]\}$.
			\RETURN $\mr{rank}(P_{n - |\mb{s}|}(\mb{c}), n - |\mb{s}|, \mc{S}' ,T') + M_1$.
			\ENDIF
			
		\end{algorithmic}
	\end{algorithm}
	
	\begin{theorem}\label{thm:ell}
	For given $n,\ell,\epsilon$ and a fixed $T$ with $n\le T\le 4n$,		the code $\widetilde{\mathcal{C}}(n,\ell,T)$ encodes a binary sequence of length $\lceil \log_2{N(n,\ell,T)}\rceil$ to a DNA sequence of length $n$  with $\ell$-RLL and synthesis time at most $T$. This results in a redundancy of
	 $r_{\ell,T}=n-\lceil \log_4{N(n,\ell,T)}\rceil$. In addition, the encoding and decoding algorithms run in $O(n^3\ell^2)$ time, thus, $\widetilde{\mathcal{C}}(n,\ell,T)$ is a capacity-achieving constrained code for DNA synthesis with polynomial-time complexity.
	\end{theorem}
	\begin{proof}
		Since we enumerate all valid DNA sequences, the code $\widetilde{\mathcal{C}}(n,\ell,T)$ is capacity-achieving and the redundancy of $\mathcal{C}$ is $r_{\ell,T}=n-\lceil \log_4{N(n,\ell,T)}\rceil$, which are straightforward.
		
		The computational complexity includes the following two parts:
		i). Precomputing and storing of $N$ (offline preprocessing):
		We need to precompute and store $N(n',\mathbf{s}',T')$ for all $n'\in[1, n], \mathbf{s}'\in\mathcal{S},T'\in[1,T]$, i.e., there are $O(n^2\ell)$ states for $N$.
		Each $N(n', \mathbf{s}', T')$ is a denary number of length $O(n)$.
		According to \eqref{eqn:state-ell}, the computation of $N(n,\mathbf{s},T)$
		is based on the other $O(\ell)$ states, indicating the computational
		complexity for computing $N$ is $O(n^3\ell^2)$.
		ii). Computation in Algorithms \ref{Unrank} and \ref{Rank} for a given $N$ (online
		processing): Both Algorithms \ref{Unrank} and \ref{Rank} iterate at most $n$
		times. In each iteration, the computation of $M_1$ dominates the
		complexity. As can be seen from Line \ref{line7} of Algorithm \ref{Unrank} and
		Line \ref{line2-8} of Algorithm \ref{Rank}, the set $\mc{S}$ is of size $O(\ell)$ for each iteration, and $N(n,\mathbf{s},T)$ is a denary
		number of length $O(n)$. Therefore, the computation of $M_1$ has
		computational complexity $O(n^2\ell)$.
	\end{proof}
	
	\subsection{$(\ell,\epsilon)$-Constrained Codes for DNA Synthesis}\label{sec:lgc-indel}
		For a fixed $T$ with $n\le T\le 4n$, denote the set of length-$n$ sequences satisfying the $(\ell,\epsilon)$-constraints with synthesis time at most $T$  by  $\widetilde{\mathcal{C}}(n,\ell,\epsilon,T)$.
		Denote all the sequences with length $n$, last run $\mathbf{s}$, GC-content $m$, and  synthesis time at most $T$ by $\mathcal{C}(n,\mathbf{s},m,T)$.
	Thus, we can write
	\begin{eqnarray*}
		\widetilde{\mathcal{C}}(n,\ell,\epsilon,T)=\bigcup_{\mathbf{s}\in\mathcal{S},m\in\mathcal{M}}\mathcal{C}(n,\mathbf{s},m,T),
	\end{eqnarray*}
where $\mathcal{S}$ and $\mathcal{M}$ are defined in \eqref{set:S} and \eqref{eqn:M}, respectively.

	Let $N(n,\mathbf{s},m,T)$ be the size of $\mathcal{C}(n,\mathbf{s},m,T)$, i.e., $N(n,\mathbf{s},m,T)=|\mathcal{C}(n,\mathbf{s},m,T)|$. In the following, we derive a recursion for computing $N(n,\mathbf{s},m,T)$.
	\begin{theorem}
		The number of  length-$n$ DNA sequences with $(\ell,\epsilon)$-constraints and synthesis time at most $T$  can be computed by 
			\begin{eqnarray*}
			&&\hspace{-2.5em}	N(n,\mathbf{s},m,T)\nonumber\\
				&=& \left\{\begin{array}{ll}
					0, \qquad \qquad \text{~if~}  n<|\mathbf{s}| \text{~or~} T<4(|\mathbf{s}|-1)+e(\mathbf{s}),\\
					\mbb{1}_{\{G(\mathbf{s})=m\}},\text{~if~}  n=|\mathbf{s}|\text{~and~}T\ge4(|\mathbf{s}|-1)+e(\mathbf{s}), \\
					\sum\limits_{\mathbf{s}'\in\mathcal{S}:e(\mathbf{s}')\ne e(\mathbf{s})} N(n-|\mathbf{s}|,\mathbf{s}',m-G(\mathbf{s}),T'),  \\
					\quad\qquad \qquad \text{if~}  n>|\mathbf{s}| \text{~and~} T\ge 4(|\mathbf{s}|-1)+e(\mathbf{s}),
				\end{array}\right.
			\end{eqnarray*}	
		where $\mathbf{s}\in \mathcal{S},T'=T-4(|\mathbf{s}|-1)-(e(\mathbf{s})-e(\mathbf{s}')\bmod 4)$.
	\end{theorem}
	\begin{proof}
		The proof of the first two equations are similar to those of Theorem \ref{thm:state-ell}, therefore, we only illustrate the last one.
		For any sequence $\mathbf{c}\in\mathcal{C}(n,\mathbf{s},m,T)$, 
		$G(P_{n-|\mathbf{s}|}(\mathbf{c}))=m-G(\mathbf{s})$ as $G(\mathbf{c})=m$, where $P_{n-|\mathbf{s}|}(\mathbf{c})$ is the prefix of $\mathbf{c}$ with length $n-|\mathbf{s}|$.
		
		Hence, we obtain
	\begin{eqnarray*}
			\mathcal{C}(n,\mathbf{s},m,T)=\{\mathbf{c}'||\mathbf{s}:
			\mathbf{c}'\in\mathcal{C}(n-|\mathbf{s}|,\mathbf{s}',m-G(\mathbf{s}),T'),\\
			\mathbf{s}'\in \mathcal{S} \text{~with~}e(\mathbf{s}')\ne e(\mathbf{s})\},
		\end{eqnarray*}
		 completing the proof.
	\end{proof}
	
	
	To order all the sequences in $\widetilde{\mathcal{C}}(n,\ell,\epsilon,T)$, we order the sequences logically in each subset $\mathcal{C}(n,\mathbf{s},m,T)$ by GC-content first, then the length and symbol of the last run.
	For any pairs of sequences $\mathbf{c}_1\in\mathcal{C}(n,\mathbf{s}_1,m_1,T_1)$ and $\mathbf{c}_2\in \mathcal{C}(n,\mathbf{s}_2,m_2,T_2)$, define $\mathbf{c}_1< \mathbf{c}_2$ if any of the following conditions are satisfied. 
	\begin{itemize}
		\item[a.] $m_1<m_2$;
		\item[b.]  $m_1=m_2$ and $|\mathbf{s}_1|<|\mathbf{s}_2|$;
		\item[c.] $m_1=m_2, |\mathbf{s}_1|=|\mathbf{s}_2|$ and $e(\mathbf{s}_1)<e(\mathbf{s}_2)$;
		\item[d.] $m_1=m_2, |\mathbf{s}_1|=|\mathbf{s}_2|,e(\mathbf{s}_1)=e(\mathbf{s}_2)$ and 
		$P_{n_1}(\mathbf{c}_1)<P_{n_2}(\mathbf{c}_2)$, where $n_1=n-|\mathbf{s}_1|,n_2=n-|\mathbf{s}_2|$.
	\end{itemize}	
	Additionally, for two pairs $(\mathbf{s}_1,m_1)$ and $(\mathbf{s}_2,m_2)$, if a-c are satisfied, we say $(\mathbf{s}_1,m_1)<(\mathbf{s}_2,m_2)$.
	
After computing the size of each subset and logically ordering all $(\ell,\epsilon)$-constrained sequences, we can design the encoding and decoding algorithms by unranking and ranking methods, denoted by $\mathrm{Enc}_{\ell,\epsilon,T}$ and $\mathrm{Dec}_{\ell,\epsilon,T}$, respectively. 	The details of $\mathrm{Enc}_{\ell,\epsilon,T}$ and $\mathrm{Dec}_{\ell,\epsilon,T}$ are similar to Algorithms \ref{Unrank} and \ref{Rank}.  In the following, we state the main differences.

In the unranking method, the objective is to find the  $M$-th smallest codeword $\mathbf{c}\in\widetilde{\mathcal{C}}(n,\ell,\epsilon,T)$, where $M\in[1,N(n,\ell,\epsilon,T)]$. To achieve this, we define the function $\mathrm{unrank}(M,n,\mc{S}\times\mc{M},T)$ as described in Algorithm \ref{Unrank} such that we can find the unique $(\mb{s}, m)\in \mc{S} \times \mc{M}$ and compute $M_1 < M \leq M_1 + N(n, \mb{s},m, T)$ in Line \ref{line7},
where 
\begin{eqnarray}\label{eqn:M1}
	M_1 = \sum_{(\mb{s}',m') \in \mc{S}' \times \{m'\}: (\mb{s}',m')< (\mb{s},m)}N(n, \mb{s}',m',T).
\end{eqnarray}
Then the solution is converted to return $\mathrm{unrank}(M-M_1,n - |\mb{s}|, \mc{S}' \times \{m'\},T')||\mathbf{s}$, where $m'=m-G(\mathbf{s})$.

In the ranking method, we determine the ordering of a given DNA sequence. This implies that there exists a unique integer $M\in [1,N(n,\ell,\epsilon,T)]$ corresponding to a DNA sequence $\mathbf{c}\in\widetilde{\mathcal{C}}(n,\ell,\epsilon,T)$.
We define the function $\mathrm{rank}(\mathbf{c},n,\mc{S}\times\mc{M},T)$ as outlined in Algorithm \ref{Rank}, which enables us to compute $M_1$ in Line \ref{line2-8}. Therein, $M_1$ is defined in \eqref{eqn:M1}, and $\mb{s},m$ are the last run and GC-content of $\mathbf{c}$, respectively. Then we only need to return $\mr{rank}(P_{n - |\mb{s}|}(\mb{c}), n - |\mb{s}|, \mc{S}'\times\{m'\} ,T') + M_1$, where $m'=m-G(\mathbf{s})$.

	\begin{theorem}\label{thm:ell,epsilon}
		For given $n,\ell,\epsilon$ and a fixed $T$ with $n\le T\le 4n$,	the code $\widetilde{\mathcal{C}}(n,\ell,\epsilon,T)$ encodes a binary sequence of length $\lceil \log_2{N(n,\ell,\epsilon,T)}\rceil$ to a DNA sequence of length $n$  with  $\ell$-RLL, $\epsilon$-balance and synthesis time at most $T$. This results in a redundancy of $r_{\ell,\epsilon,T}=n-\lceil \log_4{N(n,\ell,\epsilon,T)}\rceil$.
		 In addition, the encoding and decoding algorithms run in $O(n^4\ell^2)$ time, thus $\widetilde{\mathcal{C}}(n,\ell,\epsilon,T)$ is a capacity-achieving  constrained code for DNA synthesis with polynomial-time complexity.
	\end{theorem}
	
	\begin{proof}
		The proof is similar to that of Theorem \ref{thm:ell}, and thus is omitted.
	\end{proof}

	

	
	
	
	\section{Constrained ECCs for DNA Synthesis}\label{sec:constrained-error-code}
In this section, we construct the constrained ECCs for DNA synthesis, that not only satisfy 
 the $\ell$-RLL and $\epsilon$-balance constraints, but also are capable of correcting one indel in DNA sequences.

	For a DNA sequence $\mathbf{c}=(c_1,c_2,\dots,c_n)$, 
	the \textit{VT syndrome} and \textit{signature vector} of $\mathbf{c}$ are defined by 
	\begin{eqnarray*}
		\mathrm{VT}(\mathbf{c})=\sum_{i=1}^n i\cdot c_i,
	\end{eqnarray*}
	and
	$$s(\mathbf{c})=(s_1(\mathbf{c}),s_2(\mathbf{c}),\dots,s_{n-1}(\mathbf{c})),$$ 
	respectively, where for each $i\in[1,n)$, $	s_i(\mathbf{c})=1$ if $c_{i+1}\ge c_i$, otherwise, it is $0$.

	Denote $m=n-\lceil\log_4 n\rceil -3$. For any $\mathbf{c}=(c_1,\dots,c_m,$ $c_{m+1},\dots,c_n)\in\Sigma_4^n$, let $c_{m+4}\cdots c_n$  be the $4$-ary representation of $\mathrm{VT}(s(\mathbf{c})) \equiv a\bmod n$.
	Then,  the \textit{systematic VT code}  constructed in \cite{chrisnata2023deletion} for $4$-ary sequences is
	\begin{align*}
		\mathrm{VT}_n(a,b) = \Big\{ \mathbf{c} = (c_1,\dots,c_m,c_{m+1},\dots,c_n) \in \Sigma_4^n: & \\
		\mathrm{VT}(s(\mathbf{c})) \equiv a \bmod n, & \\
		c_{m+1} = c_{m+2} = c_m + 2 \bmod 4, & \\
		c_{m+3} = \sum_{i=1}^{m} c_i \equiv b \bmod 4 \Big\}.
	\end{align*}
	Moreover, we define $\mathrm{VT}_{n}^{-1}(\mathbf{c},a,b)\triangleq(c_1,c_2,\dots,c_m)\in\Sigma_4^m$.
	
	For a symbol $a\in\Sigma_4$, let $f(a)=5-a$.
	Denote the length-$n$ constrained ECC which satisfies  $(\ell,\epsilon)$-constraints and can correct one indel in DNA synthesis by $\widetilde{\mathcal{C}}(n,\ell,\epsilon,T,\mathcal{B}^{DI})$.
	We propose the encoding and decoding methods of the constrained ECC for DNA synthesis in Algorithms \ref{alg:indelenc} and \ref{alg:indeldec}, denoted by $\mathrm{Enc}_{T,\mathrm{indel}}$ and $\mathrm{Dec}_{T,\mathrm{indel}}$, respectively. 
	In Algorithm \ref{alg:indelenc}, the original message $\mathbf{m}\in\{0,1\}^k$ is first encoded into a constrained codeword $\mathbf{c}'\in\widetilde{\mathcal{C}}(n,\ell,\epsilon,T_1)$, which is then concatenated  with the parity symbol $\mathbf{p}$ of a systematic VT code to form the codeword  $\mathbf{c}\in \widetilde{\mathcal{C}}(n,\ell,\epsilon,T,\mathcal{B}^{DI})$. To ensure the $(\ell,\epsilon)$-constraints, additional symbols $\beta$ and $f(\cdot)$ are introduced.
	In Algorithm \ref{alg:indeldec}, the received vector $\mathbf{c}$ is first decoded by determining the length of  $\mathbf{c}$ and then applying the VT decoder. Finally, the original message is recovered by decoding the constrained code $\widetilde{\mathcal{C}}(n,\ell,\epsilon,T_1)$.
	
	\begin{algorithm}[htbp]
		
		\caption{$\mathrm{Enc}_{T,\mathrm{indel}}$: Encoder of the designed constrained ECC for DNA synthesis.}
		\label{alg:indelenc}
		\begin{algorithmic}[1]
			\REQUIRE    $n,\ell\ge 2,\epsilon,T > 0, T_1=T-4(2\lceil\log_4 n\rceil + 8),
			k=n-(r_{\ell,\epsilon,T_1} + 2\lceil\log_4 n\rceil + 8)$,
			$\mathbf{m}\in\{0,1\}^k$.
			\ENSURE     $\mathbf{c}\in\widetilde{\mathcal{C}}(n,\ell,\epsilon,T,\mathcal{B}^{DI})$.
			
			\STATE  Obtain 
			$\mathbf{c}'=\mathrm{Enc}_{\ell,\epsilon,T_1}(\mathbf{m})\in\Sigma_4^{k_1}$ by  $\mathrm{Enc}_{\ell,\epsilon,T}$ in Section \ref{sec:lgc-indel}, where $k_1=k+r_{\ell,\epsilon,T_1}$. 
			\STATE Denote $\alpha$ as the last symbol of $\mathbf{c}'$.
			
			\STATE  Let $\alpha'=\alpha+2\bmod 4, a =\mathrm{VT}(s(\mathbf{c}')) \bmod n$ and $b =\sum_{i=1}^{k_1} c'_{i} \bmod 4$.
			
			\STATE Let $\tau_1\tau_2\cdots\tau_v$ be the quaternary representation of $a$, where $v=\lceil\log_4 n\rceil$.
			
			\STATE Let $\beta$ be an arbitrary symbol in $\Sigma_4\backslash\{f(\alpha')\}$.

			\STATE Set  $\mathbf{p} =  \alpha'\alpha'f(\alpha')f(\alpha')\beta f(\beta)\tau_1f(\tau_1)\tau_2f(\tau_2)\cdots \tau_vf(\tau_v)$ $bf(b)$.
			
			\STATE Output $\mathbf{c}=\mathbf{c}'||\mathbf{p}$.

		\end{algorithmic}
	\end{algorithm}

	\begin{algorithm}[htbp]
		
		\caption{$\mathrm{Dec}_{T,\mathrm{indel}}$: Decoder of the designed constrained ECC for DNA synthesis.}
		\label{alg:indeldec}
		\begin{algorithmic}[1]
			\REQUIRE   $n,\ell\ge 2,\epsilon,T > 0, T_1=T-4(2\lceil\log_4 n\rceil + 8), \mathbf{c}\in\Sigma_4^*=\Sigma_4^{n-1}\cup \Sigma_4^n\cup \Sigma_4^{n+1}$.
			\ENSURE     $\mathbf{m}\in\{0,1\}^k$.
			
			\STATE  If $|\mathbf{c}|=n$, no indel occurred, then $\mathbf{c}'=(c_1,\dots,c_{k_1})$, where  $k_1=k+r_{\ell,\epsilon,T_1}$. \textbf{Return}  $\mathbf{m}=\mathrm{Dec}_{\ell,\epsilon,T_1}(\mathbf{c}')$.		
			\STATE If $|\mathbf{c}|=n+1$, an insertion occurred. If $c_{k_1+1}=c_{k_1+2}$, then there is no error in the data part. \textbf{Return} $\mathbf{m}=\mathrm{Dec}_{\ell,\epsilon,T_1}(c_1,\dots,c_{k_1})$.\\
			Otherwise, let  $c_{k_1+8},c_{k_1+10},\dots,c_{n-2}$ be the quaternary representation of $a$ and $b=c_{n}$. 
			\textbf{Return $\mathbf{m}=\mathrm{Dec}_{\ell,\epsilon,T_1}(\mathrm{VT}_{k_1+1}^{-1}((c_1,\dots,c_{k_1+1}),a,b))$}.
			
			\STATE If $|\mathbf{c}|=n-1$, a deletion  occurred. If $c_{k_1+1}=c_{k_1+2}$, then there is no error in the data part. \textbf{Return} $\mathbf{m}=\mathrm{Dec}_{\ell,\epsilon,T_1}(c_1,\dots,c_{k_1})$.\\
			Otherwise, let  $c_{k_1+6},c_{k_1+8},\dots,c_{n-4}$ is the quaternary representation of $a$ and $b=c_{n-2}$. 
			\textbf{Return $\mathbf{m}=\mathrm{Dec}_{\ell,\epsilon,T_1}(\mathrm{VT}_{k_1-1}^{-1}((c_1,\dots,c_{k_1-1}),a,b))$}.

		\end{algorithmic}
	\end{algorithm}
	
	\begin{theorem}\label{thm:ell,epsilon,indel}
	For given $n,\ell,\epsilon$ and a fixed $T$ with $n\le T\le 4n$,	denote $T_1=T-4(2\lceil\log_4 n\rceil + 8)$. The code $\widetilde{\mathcal{C}}(n,\ell,\epsilon,T,\mathcal{B}^{DI})$ encodes a binary sequence with length  $\lceil \log_2{N(n,\ell,\epsilon,T_1)}\rceil$ to a codeword $\mathbf{c}\in\widetilde{\mathcal{C}}(n,\ell,\epsilon,T,\mathcal{B}^{DI})$, which can correct one indel in DNA sequences. This results in a redundancy of  $r_{\ell,\epsilon,T_1} + 2\lceil\log_4 n\rceil + 8$. Additionally, the encoding and decoding algorithms run in $O(n^4\ell^2)$  time.
	\end{theorem}
	\begin{proof}
		The  properties of $(\ell,\epsilon)$-constraints, indel-correcting, along with the number of redundancy of $\widetilde{\mathcal{C}}(n,\ell,\epsilon,T,\mathcal{B}^{DI})$ have been clearly established. We now focus on the computational complexity and synthesis time.
	Since the encoder and decoder of the systematic VT codes  constructed in \cite{chrisnata2023deletion} for $4$-ary sequences  have linear time complexity, then the computational complexity of $\widetilde{\mathcal{C}}(n,\ell,\epsilon,T,\mathcal{B}^{DI})$ is the same as  that of $\widetilde{\mathcal{C}}(n,\ell,\epsilon,T_1)$, i.e., $O(n^4\ell^2)$.
		
		In addition,
		the synthesis time of  $\widetilde{\mathcal{C}}(n,\ell,\epsilon,T_1)$ is given by $T_1=T-4(2\lceil\log_4 n\rceil + 8)$. Therefore, the synthesis time of a codeword  $\mathbf{c}\in\widetilde{\mathcal{C}}(n,\ell,\epsilon,T,\mathcal{B}^{DI})$ is at most $T$, since the additional redundancy is $2\lceil\log_4 n\rceil + 8$ symbols. This completes the proof.
	\end{proof}
We compare our codes with some known results in Table \ref{tab:compare}. 
	Since our constructions in this paper are based on the enumeration technique, by analyzing the number of valid sequences, we conclude that the parameters $\gamma_{\ell},\gamma_{T},\gamma_{\ell,T}$ and $\gamma_{\ell,\epsilon, T}$ in Table \ref{tab:compare} are constants of order $O(1)$. Due to space limitations, we omit the proofs.

	\section{Conclusions}\label{sec:conclusion}
In this paper, we propose capacity-achieving constrained codes with either only $\ell$-RLL or both $\ell$-RLL and $\epsilon$-balanced constraints for DNA synthesis. Furthermore, we extend these results to design constrained ECCs that can correct a single indel for DNA synthesis. 	
Note that it is also possible to construct constrained codes with only $\epsilon$-balance, or ECCs capable of correcting an edit error  using a similar approach.

	\ifCLASSOPTIONcaptionsoff
	\newpage
	\fi

	\bibliographystyle{IEEEtran}
	\bibliography{myreference}

\end{document}